\documentclass[a4paper,11pt]{article}
\usepackage{pos}
\usepackage{yfonts}

\title{Generalized Beth-Uhlenbeck approach to the thermodynamics of quark-hadron matter}
\author*[a,b,c]{David Blaschke}
\author[d]{Oleksii Ivanytskyi}
\author[e]{Gerd R\"opke}

\affiliation[a]{Institute of Theoretical Physics, University of Wroclaw,
  plac Maxa Borna 9, 50-204 Wroclaw, Poland}

\affiliation[b]{Helmholtz-Zentrum Dresden-Rossendorf,
Bautzener Landstrasse 400, D-01328 Dresden, Germany}

\affiliation[c]{Center for Advanced Systems Understanding (CASUS),
Untermarkt 20, D-02826 G\"orlitz, Germany}

\affiliation[d]{Center for Simulations of Superdense Fluids, University of Wroclaw,
  plac Maxa Borna 9, 50-204 Wroclaw, Poland}

\affiliation[e]{Institute of Physics, University of Rostock,
Albert-Einstein Str. 23-24, D-18059 Rostock, Germany}

\emailAdd{david.blaschke@uwr.edu.pl}
\emailAdd{oleksii.ivanytskyi@uwr.edu.pl}
\emailAdd{gerd.roepke@uni-rostock.de}

\abstract{We present a unified approach to the transition from hadronic matter to quark matter where hadrons are treated as bound states of quarks which dissociate at high densities due to quark Pauli blocking. 
The newly developed approach makes use of a cluster virial expansion formulated in terms of a generalized $\Phi$-derivable approach to multi-quark correlations with bound and continuum states in their spectrum encoded in hadron phase shifts. %\cite{Blaschke:2023pqd}. 
Our model can be used to obtain thermodynamic functions not only at zero and small chemical potentials, where they are consistent with lattice QCD simulations, but also at large chemical potentials where lattice QCD simulations have the sign problem. 
By applying a reaction-kinetic criterion for the chemical freeze-out of multi-quark clusters in heavy-ion collisions, we demonstrate that the chemical freeze-out coincides with their Mott transition.
The approach can be applied to study the effects of the QCD transition on primordial black hole formation in the early Universe and on hybrid neutron star formation in supernova explosions and binary neutron star mergers.
}

\FullConference{The XVIth Quark Confinement and the Hadron Spectrum Conference (QCHSC24)\\
 19-24 August, 2024\\
 Cairns Convention Centre, Cairns, Queensland, Australia\\}

\begin{document}
\maketitle

\section{Introduction}
One of the central topics at this conference is to uncover the structure of the QCD phase diagram and its characteristic features such as the critical endpoint (CEP) in the plane of temperature ($T$) versus chemical potential ($\mu$). An excellent introduction was given in the plenary talk by Veronica Dexheimer, summarized in these Proceedings \cite{Kumar:2025mcj}. She elucidated the constrained regions, where different methods of investigating the quark-hadron transition are applicable: Lattice QCD simulations \cite{Aarts:2023vsf}, heavy-ion collisions \cite{Sorensen:2023zkk}, functional renormalization group and Dyson-Schwinger equation studies
\cite{Fischer:2018sdj,Dupuis:2020fhh,Fu:2022gou,Lu:2025cls}.
While all these methods have their merits and provide valuable insights for the investigation of the QCD phase diagram, they still have limitations which justify the development and application of effective approaches.   

One of these limitations concerns the analysis of the composition of hot, dense QCD matter in the vicinity of the hadron-to-quark matter transition. It is a challenging task to describe the hadron resonance gas as the confined phase of QCD where the hadrons are multi-quark clusters (bound and scattering states of quarks). Due to their quark substructure, the hadrons have to undergo a Mott dissociation (delocalization) transition at high phase space occupation because quark Pauli blocking entails the melting of the chiral condensate, i.e. chiral symmetry restoration.  
In this contribution we will describe the recent development of an effective Beth-Uhlenbeck approach to the cluster virial expansion of QCD matter \cite{Blaschke:2023pqd} that provides a suitable theoretical framework for a unified description of hadronic and quark-gluon plasma phases of QCD. As an application of this approach we will present the insight that the inverse Mott effect (Mott localization) for hadrons is a mechanism for their chemical freeze-out in expanding and cooling quark-gluon plasma (QGP).

\section{Generalized Beth-Uhlenbeck EOS for quark-hadron matter}

We focus our interest on the QCD transition from a QGP to hadronic matter at the pseudocritical temperature $T_c=156.5$ MeV \cite{HotQCD:2018pds} by computing EoS in the temperature range $T\in [1;1300]$ MeV.
The behavior of such a system is described through the microscopical model \cite{Blaschke:2023pqd} of a unified EoS for the thermodynamic potential of QCD, 
\begin{equation} 
\label{eq:omega_total}
    \Omega(T,\mu,\phi,\bar{\phi})=\Omega_{\rm QGP}(T,\mu,\phi,\bar{\phi})+\Omega_{\rm MHRG}(T,\mu,\phi,\bar{\phi}),
\end{equation}
which separates the QGP sector of quark and gluon quasiparticles from that of the MHRG (Mott hadron resonance gas) comprised of hadrons which are understood as quark bound states (multiquark clusters) that can undergo a Mott dissociation.

The QGP part is descriped by a Polyakov-loop Nambu--Jona-Lasinio (PNJL) model for the non-perturbative low-energy domain augmented by $\mathcal{O}(\alpha_s)$ perturbative QCD corrections,
\begin{equation}
\Omega_{\rm QGP}(T,\mu,\phi,\bar{\phi})=\Omega_{\rm PNJL}(T,\mu,\phi,\bar{\phi})+\Omega_{\rm pert}(T,\mu,\phi,\bar{\phi}).
\end{equation}
The MHRG part takes the form of a cluster decomposition of the thermodynamic potential for quark matter 
\begin{eqnarray}
\label{eq:Omega}
\Omega_{\rm MHRG}(T,\mu,\phi,\bar{\phi}) &=& \sum_{n=2}^{N}  \Omega_n(T,\mu) + \Phi\left[\left\{S_n \right\} \right],
%\nonumber
\\
\Omega_n(T,\mu) &=& c_n\left[{\rm Tr} \ln S_n^{-1} + {\rm Tr} (\Pi_n S_n) \right],
\label{eq:Omega-a}
\end{eqnarray}    
where $n$ denotes the total number of valence quarks and antiquarks in the cluster; $c_n=1/2$ for bosonic and $c_n=-1$ for fermionic clusters \cite{Vanderheyden:1998ph,Blaizot:2000fc}. 
The functional $\Phi\left[\left\{S_n \right\} \right]$ contains all two-cluster irreducible (2CI) closed-loop diagrams that can be formed with the complete set of cluster Green's functions $S_n$. 
We will restrict ourselves to a maximum number of $N=6$ quarks in the cluster and to the class of two-loop diagrams of the "sunset" type, see Ref. \cite{Blaschke:2023pqd} for details.

The density is obtained in the form of a generalized Beth-Uhlenbeck EoS
\begin{eqnarray}
\label{eq:n}
n_{\rm MHRG}(T,\mu)&=& \sum_i a_i \, d_i\, c_{a_i}\int \frac{d^3q}{(2\pi)^3} 
\int_0^\infty \frac{d\omega}{\pi}
\left\{f^{(a_i),+}_\phi 
%\right.
%\nonumber\\
%&&\left.
-\left[f^{(a_i),-}_\phi\right]^*\right\}
2 \sin ^2 \delta_{n_i}(\omega,q) \frac{\partial  \delta_{n_i}(\omega,q)}{\partial \omega} ~,
\nonumber\\
\end{eqnarray}
where the properties of the distribution function $f^{(a),+}_\phi$ and the phase shift with respect to 
reflection $\omega \to -\omega$ have been used and the 
"no sea" approximation has been employed which removes the divergent vacuum contribution.
%$d_a$ is the degeneracy factor, 
The Polyakov--loop modified distribution functions are defined as
\begin{eqnarray}
\label{eq:PL-function}
    f^{(a),\pm}_{\phi}~\stackrel{\text{(a even)}}{=}&&\frac{({\phi} - 2\bar{\phi} y_a^\pm) y^\pm_a + {y_a^\pm}^3}{1 - 3 ({\phi} - \bar{\phi} y_a^\pm) y_a^\pm - {y_a^\pm}^3}~,\\
    f^{(a),\pm}_{\phi}~\stackrel{\text{(a odd)}}{=}&&\frac{(\bar{\phi} + 2\phi y_a^\pm) y^\pm_a + {y_a^\pm}^3}{1 + 3 (\bar{\phi} + \phi y_a^\pm) y_a^\pm + {y_a^\pm}^3}~,
\end{eqnarray}
where $y^\pm_a=e^{-\left( \omega \mp a\mu\right)/T}$ 
%and $E_p=\sqrt{\vec{p}^2+M^2}$.
and $a$ is the net number of valence quarks (antiquarks) present in the cluster. 
See Ref. \cite{Blaschke:2023pqd} for a detailed derivation.

\begin{figure}[!htb]
\begin{center}
\includegraphics[width=0.32\textwidth]{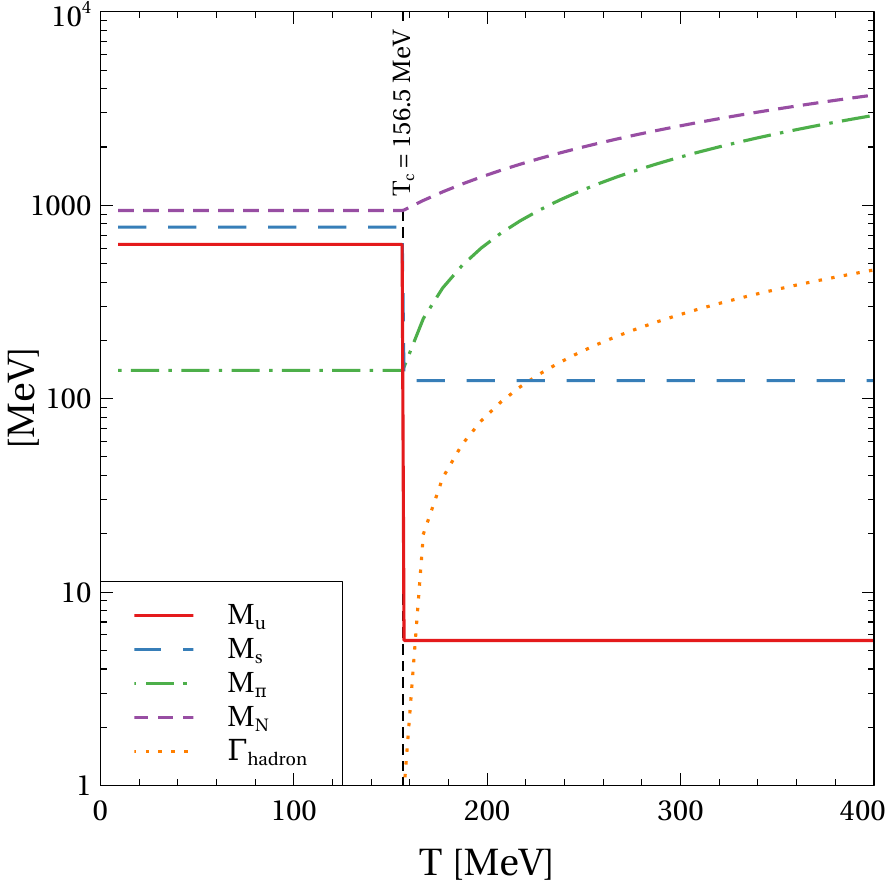}
\includegraphics[width=0.32\textwidth]{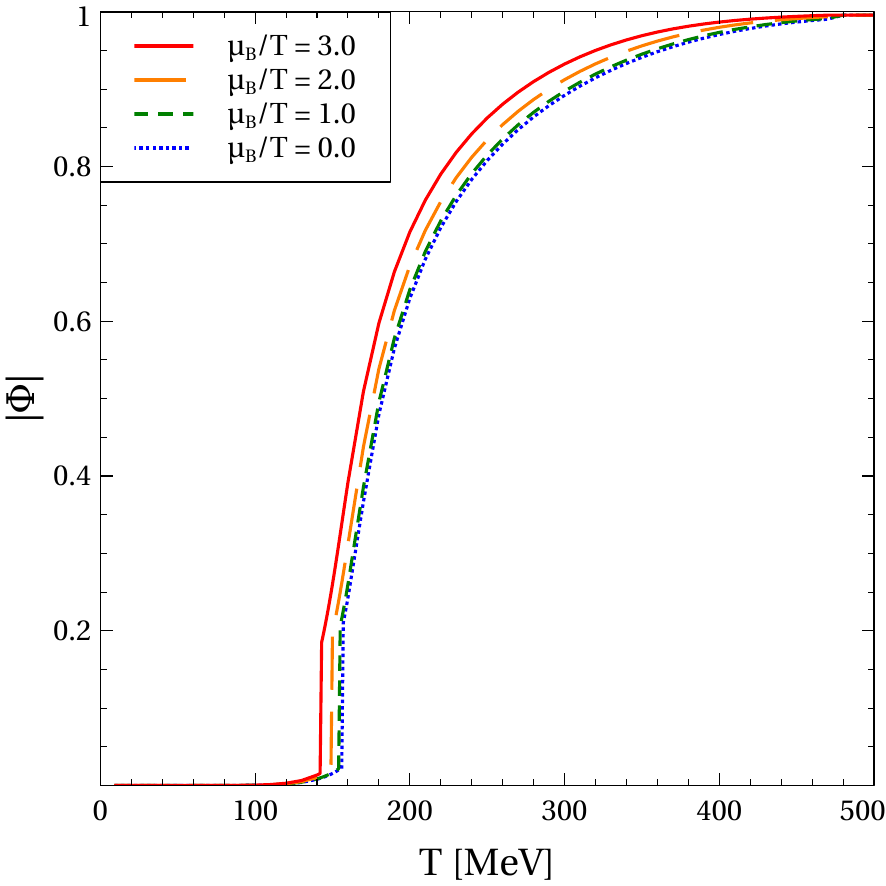}
\includegraphics[width=0.32\textwidth]{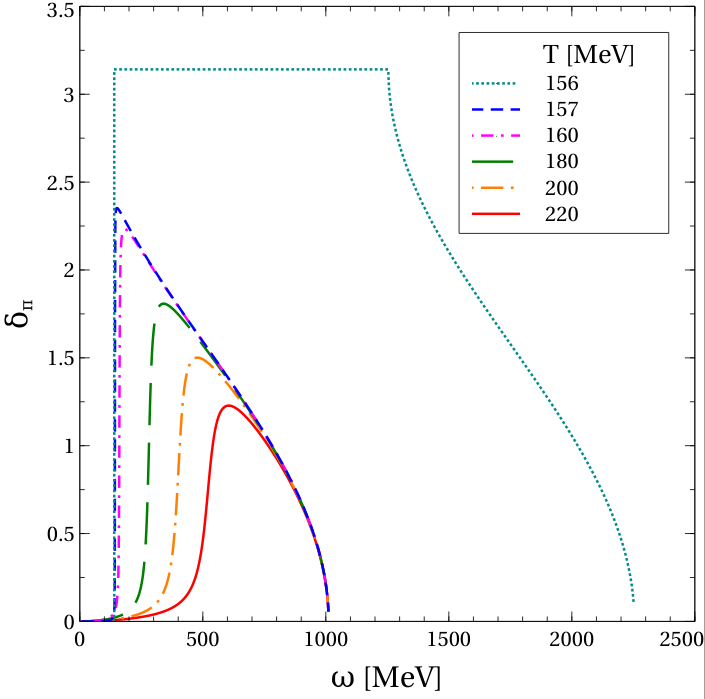}

\end{center}
\caption{Left: Mass spectrum of light and strange quarks used in this work, together with mass and decay width of the Breit-Wigner model for the pion and the nucleon as generic examples for hadrons as a function of temperature for vanishing chemical potential $\mu/T=0$.
Middle: Temperature dependence of the Polyakov loop absolute value calculated for several values of $\mu_B/T$ indicated in the legend.
Right: Step-up+continuum model for the phase shift of a hadron at rest in the medium.
}
\label{fig:masses}
\end{figure}

In an analogous manner follows for the MHRG entropy density
\begin{eqnarray}
s_{\rm MHRG}(T,\mu)&=& -\frac{\partial \Omega}{\partial T} %\nonumber\\
= \sum_i s_i(T,\mu) \nonumber\\
&=& \sum_i  d_i \, c_{a_i}\int \frac{d^3q}{(2\pi)^3}\int \frac{d\omega}{\pi}
\left\{
\sigma^{(a_i),+}_\phi 
%\right.
%\nonumber\\
%&&\left.
+\left[\sigma^{(a_i),-}_\phi\right]^* \right\}
 2 \sin ^2 \delta_{n_i}(\omega,q) \frac{\partial  \delta_{n_i}(\omega,q)}{\partial \omega} 
~,
\nonumber\\
\label{eq:s}
\end{eqnarray} 
where $\sigma^{(a)} =  f^{(a)}_\phi  \ln f^{(a)}_\phi (-)^a [1(-)^a f^{(a)}_\phi] \ln [1(-)^a f^{(a)}_\phi]$ and $f^{(a)}_\phi$ is the distribution function for a cluster with net quark number $a$ modified by the traced Polyakov loop.

\begin{figure}[t]
    \centering
    \includegraphics[width=0.48\columnwidth]{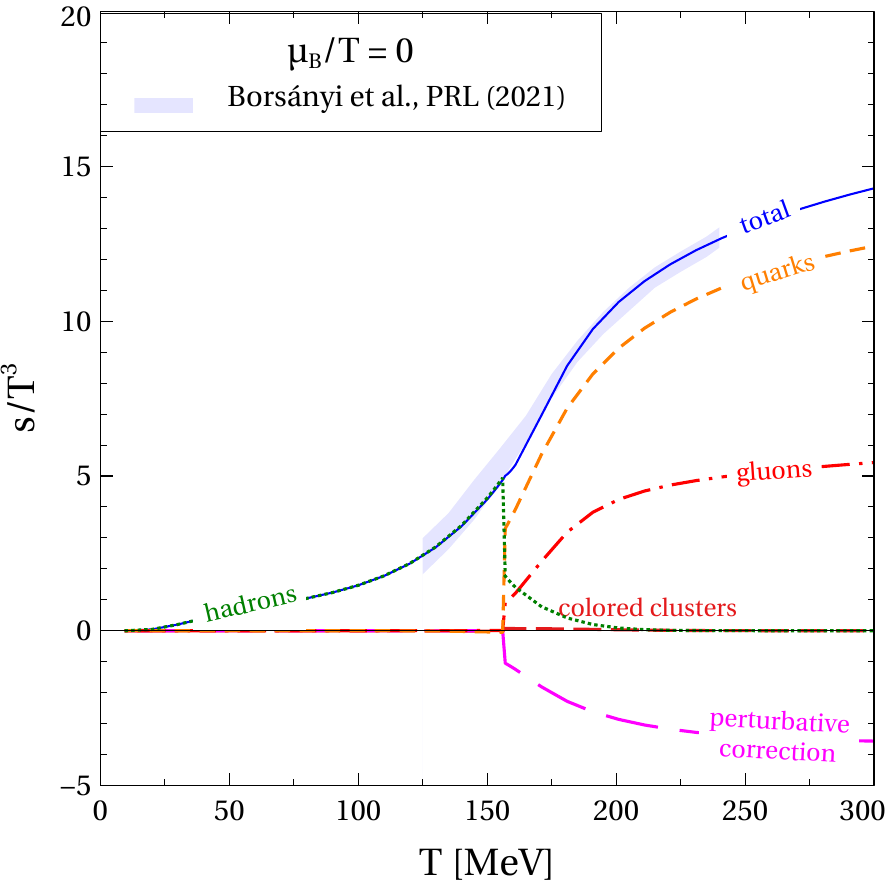}
    \includegraphics[width=0.48\columnwidth]{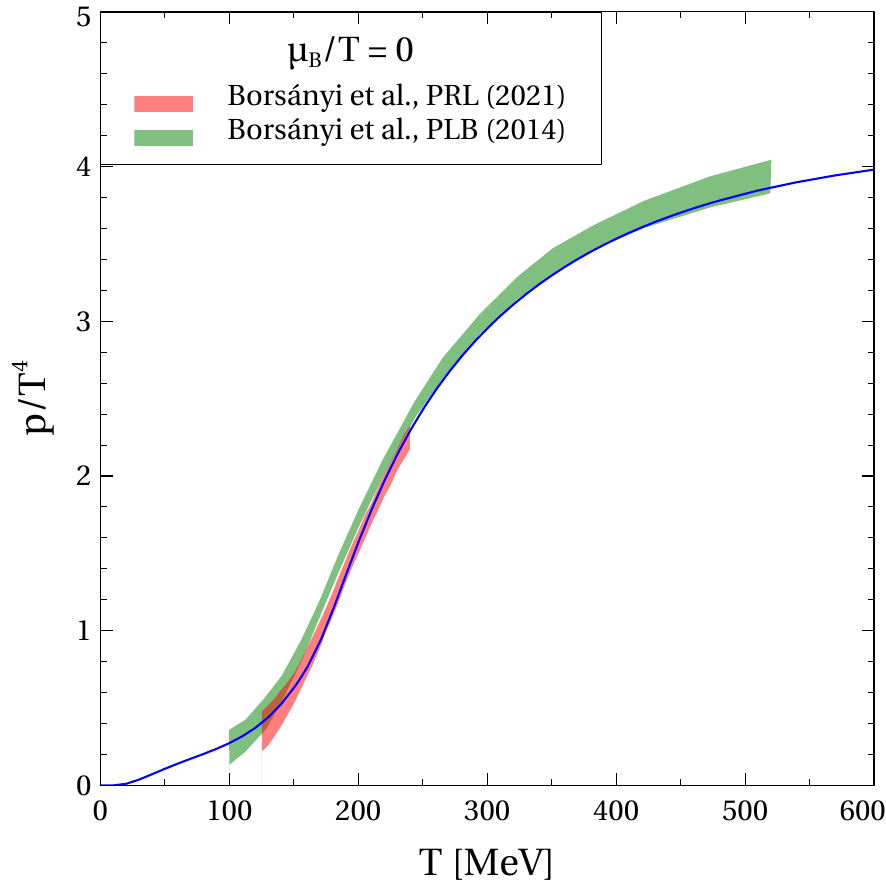}
    \caption{Left: Scaled entropy density $s/T^3$ as a function of temperature $T$ calculated at $\mu_B/T=0$ (upper panel). Partial contributions of hadrons, quarks, gluons as well as perturbative correction and total $s/T^3$ are represented by different curves indicated in figures.
    The shaded regions correspond to the lattice QCD results \cite{Borsanyi:2021sxv} in good agreement with the result of the present model (thick solid line).
    Right: Scaled pressure $p/T^4$ as a function of temperature $T$ at vanishing baryochemical potential $\mu_B/T=0$ obtained from integrating the entropy (\ref{eq:pressure}) (blue solid line), in comparison with two sets of lattice QCD data from Borsanyi et al. (2014) \cite{Borsanyi:2013bia} and Borsanyi et al. (2021) \cite{Borsanyi:2021sxv}, shown as shaded regions.
    }
    \label{fig:entropy_all}
\end{figure}

The formula for the pressure as thermodynamical potential can be obtained from Eq. (\ref{eq:n}) by integration  over the quark chemical potential $\mu$.
Analogously, it can be obtained from Eq. (\ref{eq:s}) by integration over $T$
\begin{equation}
    \label{eq:pressure}
    p(T,\mu) = \int_0^T dT' s(T',\mu)~.
\end{equation}
The resulting curve for the temperature dependence of the pressure at vanishing baryon chemical potential shows perfect agreement with the corresponding lattice QCD simulations \cite{Borsanyi:2013bia,Borsanyi:2021sxv}, see Fig. \ref{fig:entropy_all}.

\section{Chemical Freeze-out as Mott localization}

Recently, it has been demonstrated in \cite{Blaschke:2024jqd}, that the chemical freeze-out of a nuclear fireball in heavy-ion collisions with freeze-out temperatures below 100 MeV is strongly correlated with the Mott line for the alpha particles as a representative of nuclear clusters.
It is well known \cite{Andronic:2017pug} and has also been demonstrated in \cite{Blaschke:2024jqd}, that the chemical freeze-out temperature in high-energy collisions, where effects of a nonvanishing baryon chemical potential are small, coincides with the pseudocritical temperature for chiral symmetry restoration of $T_c = 156.5$ MeV found in lattice QCD simulations \cite{HotQCD:2018pds} from the peak of the chiral susceptibility.   
Here we want to demonstrate that the drop in quark masses at the chiral transition causes a Mott dissociation (Mott localization in the expanding and cooling system) of hadrons and that also in this situation the chemical freeze-out can be identified with the Mott transition. 
For an earlier formulation of this idea, see \cite{Blaschke:2011hm}.

In order to prove this claim we apply a reaction-kinetic criterion for chemical freeze-out 
of a hadron species $i$ which shall happen when the "Hubble" expansion rate of the fireball created in the heavy-ion collision $H_{\rm exp}(T)$ exceeds the reaction rate of this hadron $\tau_i^{-1}(T)$
\begin{equation}
\label{eq:CFO}
    H_{\rm exp}(T_{{\rm cf},i})=\tau_i^{-1}(T_{{\rm cf,} i}).
\end{equation}
The expansion rate can be estimated assuming the entropy conservation law $sV=const$, where the volume of the system scales with the expansion time $\tau_{\rm exp}$ as $V\propto\tau_{\rm exp}^3$.
This leads to
\begin{eqnarray}
    \label{XI}
    H_{\rm exp}=\frac{1}{\tau_{\rm exp}}=\frac{s^{1/3}}{a},
\end{eqnarray}
where $a=2.86$ \cite{Blaschke:2017lvd}.
The reaction rate is obtained within the relaxation time approximation as
\begin{eqnarray}
    \label{XII}
    \tau_i^{-1}=\sum_j\sigma_{ij} v_{\rm rel}n_j.
\end{eqnarray}
Here $v_{\rm rel}$ is the relative velocity approximated with the speed of light and 
$\sigma_{ij}$ is the total scattering cross section of two species of  hadron $i$ and $j$.
For this hadron-hadron scattering we use a geometric Povh-H\"ufner scaling law \cite{Povh:1990ad,Hufner:1992cu},
\begin{eqnarray}
    \label{XIII}
    \sigma_{ij}=\lambda\langle r_i^2\rangle\langle r_j^2\rangle,
\end{eqnarray}
where $\lambda$ is a parameter with the dimension of the string tension and $\langle r_i^2\rangle$ stands for the mean squared radius of the hadron $i$. 
The value $\lambda=0.197~\rm GeV^2$ was adjusted to describe reactions at zero temperature and baryon density and is assumed to be only weakly medium-dependent \cite{Hufner:1992cu}.
The Povh-H\"ufner law dependence of the cross section \eqref{XIII} on the size parameter of the colliding hadrons has also been found for string-flip (quark flavor exchange) reactions between heavy and light quarkonia in a study of the charmonium dissociation kinetics in a hot pion gas \cite{Martins:1994hd}.
Therefore, we expect that our assumptions are quite robust and apply also to the case of chemical (flavor) equilibration reactions in a hot hadron gas.

The cross section for scattering of hadrons on point-like quarks is estimated within the black-disc approximation,
\begin{eqnarray}
    \label{XIV}
    \sigma_{iq}=\pi\langle r_i^2\rangle.
\end{eqnarray}

\begin{figure}
    \centering
    \includegraphics[width=0.48\columnwidth]{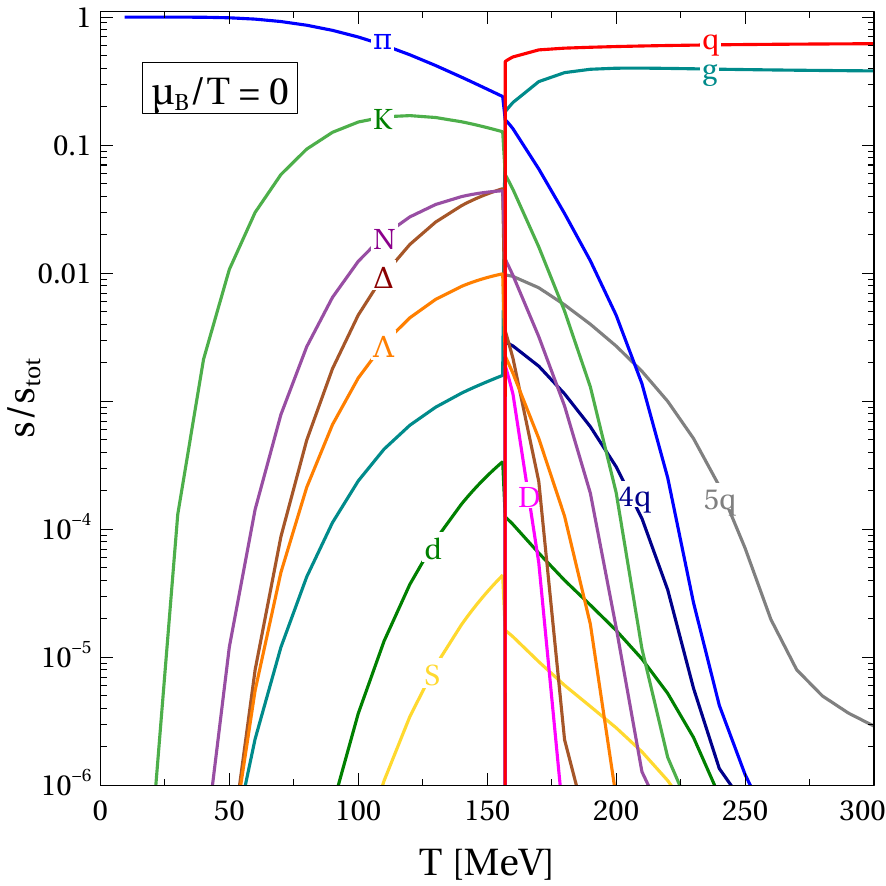}
    \includegraphics[width=0.48\linewidth]{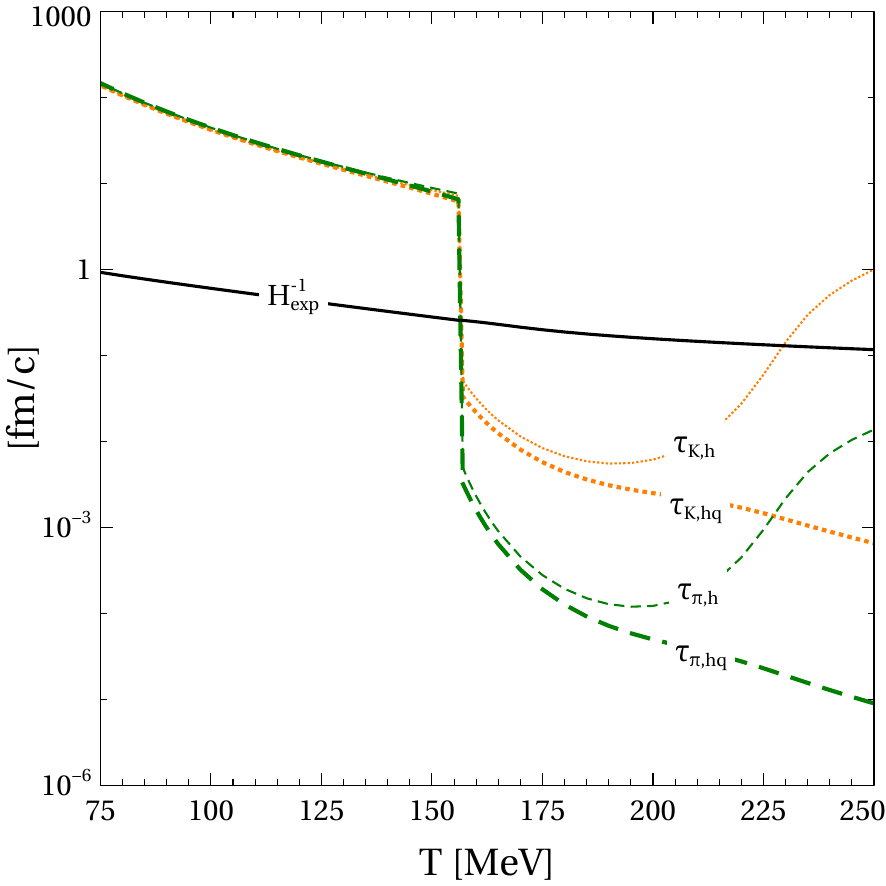}
    \caption{
    Left: Fractions of entropy density carried by different species as a function of temperature $T$ calculated at vanishing baryochemical potential. 
    Notably, the red and turquoise solid lines stands for the entropy fractions of quarks and gluons, resp.,  which are dominant in the QGP phase and strongly suppressed below the Mott transition temperature, because of the effective confinement mechanism in the present work.
    Right: Chemical freeze-out criterion. Collision times for hadron-hadron collisions according to the in-medium generalized Povh-H\"ufner law for total cross sections and (Hubble) expansion time scale.}
    \label{fig:enter-label}
\end{figure}

To proceed further, we notice that below the pseudocritical temperature most of the entropy is carried by pions and kaons, which are the lightest hadron species, which can be seen from the left panel of Fig. \ref{fig:enter-label}.
This allows us to reduce the hadron spectrum to pions and kaons.
The medium-dependent  mean squared radii of these particles obtained within the Nambu-Jona-Lasinio (NJL) model through the corresponding decay constant $f_{\pi,K}$ \cite{Hippe:1995hu}
\begin{eqnarray}
    \label{XV}
    \langle r_{\pi,K}^2\rangle=\frac{3}{4\pi^2f_{\pi,K}^2}.
\end{eqnarray}
The NJL model relates the decay constants of pions and kaons to the corresponding masses, current masses of quarks and chiral condensates of light $\langle\overline{l}l\rangle$ and strange $\langle\overline{s}s\rangle$ quarks through the Gell-Mann-Oakes-Renner relation
\begin{eqnarray}
    \label{XVI}
    M_{\pi,K}^2f_{\pi,K}^2=\frac{1}{2}
    \left(N_{\pi,K}^l\langle\overline{l}l\rangle+N_{\pi,K}^s\langle\overline{s}s\rangle\right)
    \left(N_{\pi,K}^lm_l+N_{\pi,K}^sm_s\right),
\end{eqnarray}
where $N_i^f$ denotes the number of valence (anti)quarks of flavor $f$ in hadron $i$.

The described model allows us to explain the correlation between the chemical freeze-out of hadrons and their Mott dissociation. 
The Mott dissociation of hadrons is driven by a sudden decrease of the constituent quark mass $M_f$ (see Fig. \ref{fig:masses}, left panel), which also defines the chiral condensate $\langle\overline{f}f\rangle\propto M_f$.
Thus, as seen from Eq. (\ref{XVI}), the drop in $M_f$ decreases the decay constants of pions and kaons.
This, in turn, causes a rapid growth in the mean squared radii and cross sections of hadrons, which is reflected in a discontinuous increase in reaction rate $\tau_i^{-1}$. 
Fig. \ref{fig:enter-label} shows that $\tau_i^{-1}$ jumps by 3-4 orders of magnitude, while $H_{\rm exp}$ behaves smoothly.
As a result, the chemical freeze-out condition \eqref{eq:CFO}
%$H_{\rm exp}=\tau_i^{-1}$ 
is fulfilled at the chiral restoration temperature, where the quark mass drops and this drop causes the Mott dissociation of hadrons because of the vanishing of their binding energy. 

The relation between the chiral symmetry restoration that causes the vanishing of the binding energy (Mott criterion) of a composite state and the divergence of the radius of its wave function (Mott delocalization) can be nicely demonstrated for the case of the pion. 
In their discussion of the Mott transition as "soft deconfinement", H\"ufner, Klevansky and Rehberg \cite{Hufner:1996pq} expand the pion properties in the vicinity of its Mott temperature
$T_{\rm Mott, \pi}$ and find that the pion radius diverges as
\begin{equation}
    \langle r^2\rangle_\pi^{1/2} \sim |T-T_{\rm Mott, \pi}|^{-1/2}. 
\end{equation}
Since the pion-quark coupling strength $g_{\pi qq}$ is on the one hand related to the medium-dependent pion binding energy $E_{B,\pi}(T)=|2m_q(T)-m_\pi(T)|$ as 
$g_{\pi qq}\sim E_{B,\pi}(T)^{1/4}$, but on the other to the distance to the Mott temperature as 
$g_{\pi qq}\sim |T-T_{\rm Mott, \pi}|^{1/2}$, it follows that 
\begin{equation}
    \langle r^2\rangle_\pi \sim E_{B,\pi}(T)^{-1/2}.
\end{equation}
At the Mott temperature, where the binding energy vanishes, the pion radius diverges so that with the geometric scaling of the cross section \eqref{XIII} diverges and chemical equilibrium is quickly established. Therefore, when the expanding and cooling QGP fireball passes the Mott temperature of the hadron species $i$, the corresponding hadron distribution freezes out immediately. 

\section{Conclusions and Outlook}

We have developed a systematic approach to the thermodynamics of the quark-gluon-hadron system that describes hadrons as multi-quark correlations in the form of bound states and continuum correlations within a cluster-virial expansion approach based on a cluster generalization of the $\Phi$-derivable approach. 
When a restriction to closed two-loop diagrams in cluster Green's functions is applied to the $\Phi$-functional, this approach is equivalent to the generalized Beth-Uhlenbeck approach to clustering in hot and dense Fermi systems. 
The main advantage of the approach is the microscopic description of the Mott dissociation of hadrons at finite temperatures and chemical potentials, triggered by the chiral restoration.
The complexity of color confinement in low-density quark matter is accounted for by coupling the quarks and their clusters to the Polyakov-loop background field, which serves to suppress the appearance of colored clusters in the quark confinement region.
We have compared our results for the thermodynamics of clustered quark matter with recent lattice QCD simulations at finite temperature and small chemical potentials where they are currently available and found satisfactory agreement.

One key application presented in this contribution is a microscopic justification of the coincidence of chemical freeze-out with Mott dissociation in ultra-relativistic heavy-ion collisions, which was achieved by applying a reaction kinetic freeze-out criterion and a geometric scaling of reaction cross sections with medium-dependent hadron radii.

The approach has recently been applied to the description of primordial black hole formation at the QCD transition in the early Universe \cite{Gonin:2025uvc}. It will next be extended to the situation of low temperatures and large baryon chemical potentials which is met in astrophysical systems such as neutron stars, their mergers and supernova explosions. 
For those systems an improvement of the microscopic description of the deconfinement and hadron dissociation transition is expected. The widely open question whether the transition in that region is of first order or rather a crossover transition can be attacked with a consistent microphysical approach such as the generalized Beth-Uhlenbeck approach, see \cite{Bastian:2018mmc}.

The presented approach contains some simplifying approximations, especially with respect to the 
in-medium phase shifts. A self-consistent solution of the gap equation in a correlated medium would be of interest to obtain consistent results. 
We expect that such a solution would exhibit a behaviour which justifies the assumption of a sudden switch model for the quark masses that we made in this work.
The few-quark wave equation describing the shift in binding energies and the phase shifts $\delta_i$ is subject of further investigation. In order to deal with the effects of color superconductivity, the formulation of the thermodynamic potential has to be based on Green's functions in the Nambu-Gorkov representation. 
With these future developments we will obtain consistent expressions for unified equations of state that are needed to simulate matter under extreme conditions, such as those encountered in cosmology, ultrarelativistc heavy-ion collisions and in dense astrophysical objects.

\subsection*{Acknowledgements}
D.B. and O.I. were supported by the Polish NCN
%Narodowe Centrum Nauki (NCN) 
under grant No. 2021/43/P/ST2/03319. G.R. acknowledges a joint stipend from the Alexander von Humboldt Foundation and the Foundation for Polish Science under grant No. DPN/JJL/402-4773/2022.

%\bibliographystyle{JHEP}
%\bibliography{ref}

\begin{thebibliography}{10}

\bibitem{Kumar:2025mcj}
R.~Kumar, V.~Dexheimer and J.~Jahan, \emph{{Neutron stars and Constraints for the Equation of State of Dense Matter}},  in \emph{{16th Conference on Quark Confinement and the Hadron Spectrum}}, 3, 2025 [\href{https://arxiv.org/abs/2503.23413}{{\ttfamily 2503.23413}}].

\bibitem{Aarts:2023vsf}
G.~Aarts et~al., \emph{{Phase Transitions in Particle Physics}: {Results and Perspectives from Lattice Quantum Chromo-Dynamics}}, \href{https://doi.org/10.1016/j.ppnp.2023.104070}{\emph{Prog. Part. Nucl. Phys.} {\bfseries 133} (2023) 104070} [\href{https://arxiv.org/abs/2301.04382}{{\ttfamily 2301.04382}}].

\bibitem{Sorensen:2023zkk}
A.~Sorensen et~al., \emph{{Dense nuclear matter equation of state from heavy-ion collisions}}, \href{https://doi.org/10.1016/j.ppnp.2023.104080}{\emph{Prog. Part. Nucl. Phys.} {\bfseries 134} (2024) 104080} [\href{https://arxiv.org/abs/2301.13253}{{\ttfamily 2301.13253}}].

\bibitem{Fischer:2018sdj}
C.S.~Fischer, \emph{{QCD at finite temperature and chemical potential from Dyson\textendash{}Schwinger equations}}, \href{https://doi.org/10.1016/j.ppnp.2019.01.002}{\emph{Prog. Part. Nucl. Phys.} {\bfseries 105} (2019) 1} [\href{https://arxiv.org/abs/1810.12938}{{\ttfamily 1810.12938}}].

\bibitem{Dupuis:2020fhh}
N.~Dupuis, L.~Canet, A.~Eichhorn, W.~Metzner, J.M.~Pawlowski, M.~Tissier et~al., \emph{{The nonperturbative functional renormalization group and its applications}}, \href{https://doi.org/10.1016/j.physrep.2021.01.001}{\emph{Phys. Rept.} {\bfseries 910} (2021) 1} [\href{https://arxiv.org/abs/2006.04853}{{\ttfamily 2006.04853}}].

\bibitem{Fu:2022gou}
W.-j.~Fu, \emph{{QCD at finite temperature and density within the fRG approach: an overview}}, \href{https://doi.org/10.1088/1572-9494/ac86be}{\emph{Commun. Theor. Phys.} {\bfseries 74} (2022) 097304} [\href{https://arxiv.org/abs/2205.00468}{{\ttfamily 2205.00468}}].

\bibitem{Lu:2025cls}
Y.~Lu, F.~Gao, Y.-x.~Liu and J.M.~Pawlowski, \emph{{Finite density signatures of confining and chiral dynamics in QCD thermodynamics and fluctuations of conserved charges}},  \href{https://arxiv.org/abs/2504.05099}{{\ttfamily 2504.05099}}.

\bibitem{Blaschke:2023pqd}
D.~Blaschke, M.~Cierniak, O.~Ivanytskyi and G.~R\"opke, \emph{{Thermodynamics of quark matter with multiquark clusters in an effective Beth-Uhlenbeck type approach}}, \href{https://doi.org/10.1140/epja/s10050-023-01229-8}{\emph{Eur. Phys. J. A} {\bfseries 60} (2024) 14} [\href{https://arxiv.org/abs/2308.07950}{{\ttfamily 2308.07950}}].

\bibitem{HotQCD:2018pds}
{\scshape HotQCD} collaboration, \emph{{Chiral crossover in QCD at zero and non-zero chemical potentials}}, \href{https://doi.org/10.1016/j.physletb.2019.05.013}{\emph{Phys. Lett. B} {\bfseries 795} (2019) 15} [\href{https://arxiv.org/abs/1812.08235}{{\ttfamily 1812.08235}}].

\bibitem{Vanderheyden:1998ph}
B.~Vanderheyden and G.~Baym, \emph{{Selfconsistent approximations in relativistic plasmas: Quasiparticle analysis of the thermodynamic properties}}, \href{https://doi.org/10.1023/B:JOSS.0000033166.37520.ae}{\emph{J. Statist. Phys.} {\bfseries 93} (1998) 843} [\href{https://arxiv.org/abs/hep-ph/9803300}{{\ttfamily hep-ph/9803300}}].

\bibitem{Blaizot:2000fc}
J.P.~Blaizot, E.~Iancu and A.~Rebhan, \emph{{Approximately selfconsistent resummations for the thermodynamics of the quark gluon plasma. 1. Entropy and density}}, \href{https://doi.org/10.1103/PhysRevD.63.065003}{\emph{Phys. Rev. D} {\bfseries 63} (2001) 065003} [\href{https://arxiv.org/abs/hep-ph/0005003}{{\ttfamily hep-ph/0005003}}].

\bibitem{Borsanyi:2021sxv}
S.~Bors\'anyi, Z.~Fodor, J.N.~Guenther, R.~Kara, S.D.~Katz, P.~Parotto et~al., \emph{{Lattice QCD equation of state at finite chemical potential from an alternative expansion scheme}}, \href{https://doi.org/10.1103/PhysRevLett.126.232001}{\emph{Phys. Rev. Lett.} {\bfseries 126} (2021) 232001} [\href{https://arxiv.org/abs/2102.06660}{{\ttfamily 2102.06660}}].

\bibitem{Borsanyi:2013bia}
S.~Borsanyi, Z.~Fodor, C.~Hoelbling, S.D.~Katz, S.~Krieg and K.K.~Szabo, \emph{{Full result for the QCD equation of state with 2+1 flavors}}, \href{https://doi.org/10.1016/j.physletb.2014.01.007}{\emph{Phys. Lett. B} {\bfseries 730} (2014) 99} [\href{https://arxiv.org/abs/1309.5258}{{\ttfamily 1309.5258}}].

\bibitem{Blaschke:2024jqd}
D.~Blaschke, S.~Liebing, G.~R\"opke and B.~D\"onigus, \emph{{Cluster production and the chemical freeze-out in expanding hot dense matter}}, \href{https://doi.org/10.1016/j.physletb.2024.139206}{\emph{Phys. Lett. B} {\bfseries 860} (2025) 139206} [\href{https://arxiv.org/abs/2408.01399}{{\ttfamily 2408.01399}}].

\bibitem{Andronic:2017pug}
A.~Andronic, P.~Braun-Munzinger, K.~Redlich and J.~Stachel, \emph{{Decoding the phase structure of QCD via particle production at high energy}}, \href{https://doi.org/10.1038/s41586-018-0491-6}{\emph{Nature} {\bfseries 561} (2018) 321} [\href{https://arxiv.org/abs/1710.09425}{{\ttfamily 1710.09425}}].

\bibitem{Blaschke:2011hm}
D.~Blaschke, J.~Berdermann, J.~Cleymans and K.~Redlich, \emph{{Chiral condensate and Mott-Anderson freeze-out}}, \href{https://doi.org/10.1007/s00601-011-0261-6}{\emph{Few Body Syst.} {\bfseries 53} (2012) 99} [\href{https://arxiv.org/abs/1109.5391}{{\ttfamily 1109.5391}}].

\bibitem{Blaschke:2017lvd}
D.~Blaschke, J.~Jankowski and M.~Naskret, \emph{{Formation of hadrons at chemical freeze-out}},  \href{https://arxiv.org/abs/1705.00169}{{\ttfamily 1705.00169}}.

\bibitem{Povh:1990ad}
B.~Povh and J.~H{\"u}fner, \emph{{Systematics of strong interaction radii for hadrons}}, \href{https://doi.org/10.1016/0370-2693(90)90707-D}{\emph{Phys. Lett. B} {\bfseries 245} (1990) 653}.

\bibitem{Hufner:1992cu}
J.~H{\"u}fner and B.~Povh, \emph{{Diffractive elastic scattering and hadronic radii: Geometric and pomeron approaches}}, \href{https://doi.org/10.1103/PhysRevD.46.990}{\emph{Phys. Rev. D} {\bfseries 46} (1992) 990}.

\bibitem{Martins:1994hd}
K.~Martins, D.~Blaschke and E.~Quack, \emph{{Quark exchange model for charmonium dissociation in hot hadronic matter}}, \href{https://doi.org/10.1103/PhysRevC.51.2723}{\emph{Phys. Rev. C} {\bfseries 51} (1995) 2723} [\href{https://arxiv.org/abs/hep-ph/9411302}{{\ttfamily hep-ph/9411302}}].

\bibitem{Hippe:1995hu}
H.J.~Hippe and S.P.~Klevansky, \emph{{Nambu-Jona-Lasinio model compared with chiral perturbation theory: The Pion radius in SU(2) revisited}}, \href{https://doi.org/10.1103/PhysRevC.52.2172}{\emph{Phys. Rev. C} {\bfseries 52} (1995) 2172}.

\bibitem{Hufner:1996pq}
J.~H{\"u}fner, S.P.~Klevansky and P.~Rehberg, \emph{{Soft deconfinement - critical phenomena at the Mott transition in a field theory for quarks and mesons}}, \href{https://doi.org/10.1016/0375-9474(96)00206-0}{\emph{Nucl. Phys. A} {\bfseries 606} (1996) 260}.

\bibitem{Gonin:2025uvc}
M.~Gonin, G.~Hasinger, D.~Blaschke, O.~Ivanytskyi and G.~R\"opke, \emph{{Primordial black-hole formation and heavy r-process element synthesis from the cosmological QCD transition. Two aspects of an inhomogeneous early Universe}}, {\emph{Eur. Phys. J. A} {\bfseries 61} (2025) in production} [\href{https://arxiv.org/abs/2505.05463}{{\ttfamily 2505.05463}}].

\bibitem{Bastian:2018mmc}
N.-U.F.~Bastian and D.B.~Blaschke, \emph{{A unified quark-nuclear matter equation of state from the cluster virial expansion within the generalized Beth\textendash{}Uhlenbeck approach}}, \href{https://doi.org/10.1140/epja/s10050-021-00343-9}{\emph{Eur. Phys. J. A} {\bfseries 57} (2021) 35} [\href{https://arxiv.org/abs/1812.11766}{{\ttfamily 1812.11766}}].

\end{thebibliography}

\providecommand{\href}[2]{#2}\begingroup\raggedright\endgroup

\end{document}